\newcommand{\be}{\begin{equation}}
\newcommand{\ee}{\end{equation}}
\newcommand{\bea}{\begin{eqnarray}}
\newcommand{\eea}{\end{eqnarray}}

\documentclass[a4paper,11pt,onecolumn]{article}
\pdfoutput=1
\usepackage{jheppub}
\usepackage[T1]{fontenc}
\usepackage{graphicx}
\usepackage{dcolumn}
\usepackage{bm}
\usepackage{amsmath}
\usepackage{amssymb}
\usepackage{slashed}
\usepackage{accents}
\def\d{d\kern-.8 ex\vrule height 1.3 ex depth-1.24 ex width .7 ex \kern .15 ex}
\def\D{D\kern-1.7 ex\vrule height .87 ex depth-.8 ex width .7 ex \kern .95 ex}

\title{Correlation functions for open strings and chaos}

\author[a,b]{Vladan {\D}uki\'c}
\author[b]{and Mihailo \v{C}ubrovi\'c}
\affiliation[a]{Faculty of Physics, University of Belgrade, Studentski Trg 12-16, 11000 Belgrade, Serbia}
\affiliation[b]{Center for the Study of Complex Systems, Institute of Physics Belgrade, University of Belgrade, Pregrevica 118, 11080 Belgrade, Serbia}
\emailAdd{djukic@ipb.ac.rs}
\emailAdd{cubrovic@ipb.ac.rs}

\date{\today}

\abstract{We study the holographic interpretation of the bulk instability, i.e. the bulk Lyapunov exponent in the motion of open classical bosonic strings in AdS black hole/brane/ string backgrounds. In the vicinity of homogeneous and isotropic horizons the bulk Lyapunov exponent saturates the MSS chaos bound but in fact has nothing to do with chaos as our string configurations live in an integrable sector. In the D1-D5-p black string background, the bulk Lyapunov exponent is deformed away from the MSS value both by the rotation (the infrared deformation) and the existence of an asymptotically flat region (the ultraviolet deformation). The dynamics is still integrable and has nothing to do with chaos (either in gravity or in field theory). Instead, the bulk Lyapunov scale captures the imaginary part of quasinormal mode frequencies. Therefore, the meaning of the bulk chaos is that it determines the thermal decay rate due to the coupling to the heat bath, i.e. the horizon.}

\begin{document} 
\maketitle
\flushbottom

\section{Introduction}\label{secintro}

Chaos in string theory has traversed the way from an arcane and little-noticed topic to a mainstream field, thanks to the ideas of fast scrambling and black holes as the fastest scramblers in nature \cite{FastScramblers}, the Maldacena-Shenker-Stanford (MSS) maximum chaos bound for strongly coupled field theories with black hole duals \cite{MSSbound} and the notion of out-of-time ordered correlators (OTOC) \cite{BlackButter,ShocksLocal,ShocksStringy} and their applications in the physics of chaotic strongly coupled systems \cite{MaldacenaSYK,VandorenSYK,Garcia-Garcia:2017bkg}. An important motor of the field is also the connection to recent progress on the black hole information problem \cite{FastScramblersHayden,BHinfoRMP,BHinfoPenington:2019,BHinfoAlmheiri:2019} and the related puzzle of factorization \cite{StanfordNoiseWH,SaadWH,SaadWHfact,Pollack:2020gfa,Mukhametzhanov:2021hdi,BanerjeeGeomPhases,Cubrovic:2022pdh,Kruthoff}. In \cite{DeBoerVeghStringBnd} it was demonstrated for the first time that the MSS scale characterizes also the time-disordered correlation functions on a string worldsheet, provided that the induced metric has a horizon and thus mimics black hole physics. The guiding idea through all these topics is of course the AdS/CFT duality, the unifying principle of many topics in string theory and gravity. Our primary interest thus lies in the dynamics in asymptotically AdS backgrounds.

Among the many questions which have opened up, there is one seemingly technical but in fact physically important subtlety. Several papers have reported the saturation of the bound $2\pi T$ for \emph{bulk} orbits of particles \cite{Hashimoto:2016dfz,Dalui:2020qpt}, or its slight modification/generalization for fields \cite{Hashimoto:2016dfz} and strings \cite{Nunez:2018ags,Nunez:2018qcj,Cubrovic:2019qee,Ma:2019ewq,Roychowdhury:2021jqt,Yu:2023spr,Dutta:2023yhx}; the systematic answer to the question of the bulk Lyapunov exponent is given in \cite{Giataganas:2021ghs}. However, a very simple question arises: \emph{why should there ever be an MSS-like bound for bulk Lyapunov exponents?} The OTOC exponent and its MSS bound $\lambda=2\pi T$ in principle have no simple relation to the classical bulk motion and its Lyapunov exponent: the former is a property of a time-dependent correlation function in dual CFT, determined by a 4-wave scattering amplitude in the bulk, and the latter is the solution of a bulk equation of motion, for a single orbit, with no scattering and thus no OTOC-like interpretation in the bulk. This relates to a more general question: what is the CFT dual of a bulk orbit (and its Lyapunov instability exponent)? Some important work was done on this issue \cite{Dray:1984ha,Dray:1985yt,Kraus:1998hv,Balasubramanian:particle,Louko:particle,Russo:2002sr}, and the outcome is that a bulk particle is dual to a shock wave perturbation of the dual CFT. But many details are still missing; in particular, the answer cited above holds for a geodesic with both endpoints on the AdS boundary; it is less clear what the CFT dual is for an orbit not reaching the boundary. 

Paradoxically, a string in the bulk, specifically an open string, is perhaps an easier case for study. It is long known that a static or dragging string, with one endpoint in the interior and the other on the boundary, is dual to a heavy quark in the quark-gluon plasma of the supersymmetric Yang-Mills gauge theory \cite{HerzogDragForce,GubserDragForce}. Likewise, an open string with both endpoints on the boundary represents a quark-antiquark pair \cite{Maldacena:1998im,Rey:1998bq,Brandhuber:1998bs,Rey:1998ik,Hashimoto:2018fkb}, and encodes information on the confinement mechanism. It is thus a convenient framework to pose our main question: \emph{what is the meaning of the bulk Lyapunov exponent and what does it have to do with the MSS bound?}

In this work we give a partial answer to the question and demonstrate it by a number of case studies involving bosonic open strings in various backgrounds.\footnote{While the dynamics of a superstring would be an interesting problem to study, in this work we stick solely to the bosonic sector. This is enough to understand the principles, and also to model holographically the dynamics of a heavy quark in Yang-Mills plasma.} There is, in fact, no unique answer to the question of the CFT dual to a bulk Lyapunov exponent: just as various string configurations have various field theory duals (a quark, a bound pair of quarks, an EPR pair, an accelerating quark...), likewise the Lyapunov stability of these different solutions will have different meanings. Furthermore, on the string worldsheet there are two coordinates thus we have two Lyapunov exponents, with different CFT meanings.

We also find that the MSS form of the exponent is really a red herring: in the strict infinite-coupling, infinite-$N$ limit and with maximal symmetry, $2\pi T$ becomes a natural scale which has to appear in all fluctuation equations. As soon as we decrease symmetry (e.g. by considering a D1-D5-p bound state in the bulk that breaks rotational invariance) or include stringy effects, the bulk exponent (as well as OTOC \cite{ShocksStringy} and other CFT correlation functions) undergo corrections, and do not coincide anymore (neither among themselves nor with the MSS bound). Recent work on universal near-horizon symmetries \cite{Lin:2019qwu,fake:geometry} has shone additional light on the issue, allowing us to view the MSS scale as the fundamental property of black hole horizons, so it can appear in any CFT correlator which is sensitive to temperature $T$, i.e. which probes the energy scales smaller than $T$. The puzzle of "why $2\pi T$ pops out everywhere" is thus a fake issue: it disappears as soon as leading corrections or broken symmetries are taken into account.

The sharpest finding of our analysis is that the bulk Lyapunov exponents of a probe string in fact reproduce the quasi-normal mode (QNM) spectrum of the black hole or black string background; in other words, they correspond to the thermal decay rate. Since this rate is also determined by the temperature times an $O(1)$ factor, we feel that this also provides an explanation for the origin of an MSS-like expression in bulk dynamics. It also clearly spells out that the relation of bulk instability scale to OTOC-ology and chaos in dual CFT is fake. In terms of the relation to QNM, our work is a stringy generalization of a similar result for geodesics \cite{Cardoso:2008bp,Bianchi:2020des} and scalar waves \cite{Hintz:2015jkj, Motl:2003cd}.

The plan of the paper is the following. In Section \ref{secadss} we study the simplest possible case: open string in AdS-Schwarzschild background, where the basic message already appears -- the bulk Lyapunov exponent is $2\pi T$ but it is not related to chaos. In Section \ref{secother} we demonstrate the same findings on more general background. In Section \ref{secd1d5p} we address the dynamics of the string probe in the D1-D5-p and related backgrounds. Here we do a more detailed study, comparing the bulk variational equations to the retarded correlators dual to the the string fluctuations, and finding that the bulk Lyapunov instability really described the quasinormal modes and hence the thermal decay rate in field theory. The final section sums up the conclusions.

\section{Open string in AdS-Schwarzschild background}\label{secadss}

Our goal is to study the linear stability and fluctuations of classical solutions for the static open string stretching from the boundary to the horizon of an AdS black hole, the well-known simple holographic probe for a heavy quark in quark-gluon plasma.\footnote{In fact, to be precise, a heavy colored particle in super-Yang-Mills plasma.} Therefore, we write down the string action, derive the equations of motion and variational equations. Throughout the paper we consider only the bosonic sector of the string. Most of the time we will use the Polyakov action, but sometimes we will switch to the Nambu-Goto action, depending on the problem at hand.

\subsection{Setup and radial fluctuations}\label{subsecadsssetup}

Dynamics of a string in $D+1$-dimensional AdS-Schwarzschild spacetime with the time coordinate $t$, radial coordinate $r$, transverse spatial coordinates $x_i$ ($i=1,\ldots D-1$) and the horizon at $r_h$:
\be
    \label{adss3}
    ds^2 \equiv G_{\mu \nu}(x)dx^{\mu}dx^{\nu} = r^2 \left( -h(r) dt^2 + d\vec{x}^2 \right)+\frac{dr^2}{r^2 h(r)}, \quad h(r) = 1-\left( \frac{r_h}{r} \right)^D,
\ee
can be described by the Polyakov action for the string:
\be
\label{poly}
S_\mathrm{P}=-\frac{1}{2\pi\alpha'} \int d\tau d\sigma \ {} \eta^{\alpha \beta} \partial_{\alpha}X^{\mu}\partial_{\beta}X^{\nu} G_{\mu \nu}(X). 
\ee
Here and in the rest of the paper $\alpha, \beta, \cdots \in \{ \tau, \sigma\}$ and $\mu, \nu ,\cdots \in \{ t, r, \vec{x} \}$ stand for worldsheet and spacetime indices respectively. Latin indices $i,j, \cdots$ count the transverse coordinates $x_1,\ldots x_{D-2}$. As we know \cite{HerzogDragForce}, the equations of motion are consistent with the following ansatz:
\be
\label{ansatzpolyak}
t=t(\tau), \quad R = R(\sigma), \quad  X_1 = X_1(\tau,\sigma), \quad X_j=X_j(\tau),~ j=2, \dots, D-1,
\ee
describing a string stretching from the horizon at $r_h$ to the boundary at $r=\infty$. It is easiest to impose the flat worldsheet metric and solve for the Virasoro constraints together with the equations of motion:
\begin{eqnarray}
&& \ddot t(\tau) = 0, \quad  \ddot X_j(\tau) = 0, ~ j=2, \dots, D-1 \label{eomtxj} \\
\nonumber
&&- 2 h^3(R)R^4(\sigma) - \left( R(\sigma)h'(R) + 2 h(R) \right){R'}^2(\sigma) + 2 h(R)R(\sigma)R''(\sigma) + \\
&&+ h^2(R)R^4(\sigma) \left[ -R(\sigma) h'(R)+2 \left ({X'_1}^2(\tau,\sigma)+{\dot{X}_1}^2(\tau,\sigma) +\sum_{j=2}^{D-1}{\dot X_j}^2(\tau) \right) \right] = 0, \label{eomr}\\ 
&& 2 R'(\sigma)X'_1(\tau, \sigma) + R(\sigma) \left( X''_1(\tau, \sigma) - \ddot X_1(\tau,\sigma) \right) = 0, \label{eomx1}\\
&&\frac{{R'}^2(\sigma)}{R^4(\sigma)}+h(R) \left( - h(R) {\dot t}^2(\tau) + {X'_1}^2(\tau, \sigma)+{\dot{X}_1}^2(\tau,\sigma) + \sum_{j=2}^{D-1}{\dot X_j}^2(\tau) \right) = 0, \label{vir}\\
&&\label{vir22}X_1' \cdot \dot X_1=0.
\end{eqnarray}
The equations for $t$, $X_2, \dots, X_{D-1}$ (\ref{eomtxj})  are trivially satisfied when these are functions linear in $\tau$, thus we can set $t=\tau$ and $X_j = \mathrm{const}$. Moreover, the second constraint~(\ref{vir22}) requires $X_1$ to depend on one variable only. For now we choose $X_1=X_1(\sigma)$, i.e. the static open string/heavy quark (later on we will study both space- and time-dependent fluctuations). Now the remaining equation for $R(\sigma)$ (Eq.~\ref{eomr}), together with the nontrivial Virasoro constraint~(\ref{vir}), also decouples from $X_1(\sigma)$ and simplifies to the following form
\begin{equation}
\label{eomZ}
4h^3(R)R^3(\sigma)+h^2(R)h'(R)R^4(\sigma)+ h'(R){R'}^2(\sigma)- 2 h(R)R''(\sigma) = 0.
\end{equation}
The same equation can be derived from the effective Lagrangian, obtained by first substituting the trivial solutions $t=\tau$ and $X_j = 0, ~ \forall j\neq 1$ into the Polyakov Lagrangian, and then making use of the Virasoro constraint~(\ref{vir}) to eliminate ${X_1'}^2$:
\be
\label{lag1}\mathcal{L}_\mathrm{eff} = \frac{ - h^2(R) - f(R){R'}^2(\sigma) - h(R){X_1'}^2(\sigma) }{\sqrt{f(R)}h(R)}.
\ee
This Lagrangian has the worldsheet energy as its integral of motion and is thus integrable, as we will argue more rigorously in the following section. 

\subsubsection{Variational equations and Lyapunov exponents}\label{subsubsecadsslyap}

We will now write down the variational equation corresponding to the on-shell equation of motion (\ref{eomZ}). We will find that the solution to the variational equation is an exponential function, which defines the Lyapunov exponent the usual way. The unusual detail is the fact that both $R(\sigma)$ and its variation $\delta R(\sigma)$ depend on the \emph{spatial} coordinate $\sigma$. Studying the spatial dependence of the worldsheet field and calling it dynamics as we do might be controversial; so is the term Lyapunov exponent for the growth exponent of the variation $\delta R(\sigma)$. The important difference between $\sigma$ and $\tau$ dynamics is that the worldsheet time is unbounded and one can define asymptotic quantities as it is usually done for the Lyapunov exponent (defining it as the limit of small initial variation and long-time evolution $\lambda=\lim_{t\to\infty}\lim_{\delta x(0)\to 0}\log\left(\delta x\left(t\right)\right)/t$, for some generic coordinate $x$). The extent of the $\sigma$ coordinate is finite and there is no analogue to $\lim_t\to\infty$. Therefore, while we talk all the time of Lyapunov exponents, we really study what is often called finite-time Lyapunov indicator in the context of time evolution, i.e. the exponent defined locally rather than asymptotically. This is however often assumed as a matter of course: the bulk Lyapunov exponent (in time) computed, e.g. in \cite{Hashimoto:2016dfz,Cubrovic:2019qee,Cardoso:2008bp}, is also the finite-time quantity.

We are mainly interested in studying the variation near the horizon. To that end, we substitute $R(\sigma) \mapsto r_0 + \varepsilon \delta R(\sigma)$ into Eq.~(\ref{eomZ}), expand it in $\varepsilon$ small to linear order, and finally take the limit $r_0\to r_h$.\footnote{The order of limits is important because of the coordinate singularity at the horizon. If we immediately put $R(\sigma) \mapsto r_h + \varepsilon \delta R(\sigma)$ we end up with an equivalent variational equation which is however \emph{nonlinear}: $\delta R''-\delta R'^2/(2\delta R)-(D^2/2)r_h^2\delta R=0$; the nonlinearity stems from the vanishing of a term in the on-shell equation at the horizon, so that a linear expansion in $\varepsilon$ yields an equation which is not necessarily linear in $\delta R$ and its derivatives. If we try to expand the Lagrangian $\mathcal{L}_\mathrm{eff}$ and then solve the equation which follows from the leading correction to $\mathcal{L}_\mathrm{eff}$ (quadratic in $\varepsilon$), the coefficients of this equation diverge at $r=r_h$. Again, all of this is merely about the coordinate system we use; we could work in Kruskal-Szekeres coordinates and avoid the problem.} This yields the near-horizon variational equation:
\be
\delta R'' - D^2r_h^2 \delta R = 0.\label{vareqadss}
\ee
The solution is thus $\delta R \propto e^{\pm 2 \lambda_L \sigma}$ with a pair of Lyapunov exponents of equal magnitude and with opposite signs, as it has to happen in a Hamiltonian system. The exponent formally coincides with the MSS bound:
\be
\lambda_L = \frac{Dr_h}{2} = 2 \pi T.\label{2pit}
\ee
The reason why we define the Lyapunov exponent with a factor of $2$, i.e. through $\delta R \propto e^{\pm 2 \lambda_L \sigma}$ instead of $\delta R \propto e^{\pm\lambda_L \sigma}$ is that the same expression appears also in the OTOC growth, and follows from the definition of OTOC on the thermal circle. Here, for bulk equations of motion, this argument is irrelevant (these are different quantities!) but we nevertheless want to stay consistent with the definition of the MSS bound as we want to compare the two situations.

We have shown that the exponent is $2\pi T$ regardless of the spacetime dimension or any other parameters save the temperature. While it is tempting to call this "maximal chaos in the bulk", we will soon show that this system is not chaotic at all. Therefore we should interpret $\lambda_L$ not as a measure of chaos (neither in the bulk nor in CFT) but as some characteristic scale related to the near-horizon physics, that will likely correspond to relaxation time of some perturbations around a thermal horizon. Toward the end of the paper we will make this precise.

\subsection{Integrability of the static open string}\label{subsecadssinteg}

Here we prove that the simplest open string configuration -- a static open string at the horizon of the AdS-Schwarzschild geometry -- is integrable, unlike the motion of a closed winding string which is nonintegrable in the presence of a black hole \cite{StringsNonIntTseyt,BasuAdSS}. We emphasize that this does \emph{not} imply integrability of the open string for generic boundary conditions.

We will preform the same type of analysis that is done in \cite{StringsNonIntTseyt}, exploiting the Kovacic algorithm \cite{Korepin:1993kvr,ruiz1999differential,StringsNonIntCherv,StringsNonIntTseyt,BasuAnalytic}. The algorithm can be described as follows: (1) find a family of solutions to the equations of motion defining an invariant plane in the parameter space (2) write down the normal variational equation (NVE) for the invariant plane (3) solve the NVE so obtained and check whether it is expressible in terms of Liouvillian functions; these are the elementary functions (powers, exponentials, trigonometric functions and their inverses), rational functions of such elementary functions, and their integrals.\footnote{One can find the reasoning behind this algorithm in the literature. In brief, the existence of such a solution is equivalent to the solvability of the identity component $G^0$ of the Galois group; conversely, their nonexistence is equivalent to $G^0$ being not solvable, and hence non-Abelian. Non-Abelian nature of $G^0$ tells us that no complete system of integrals of motion in involution exists, therefore the system is nonintegrable.}

We want to show the integrability of the system described by the effective Lagrangian in Eq.~(\ref{lag1}). One obvious invariant plane is the $R-X_1$ plane for a straight string solution:
\be
R(\sigma) = r_h,~ X_1(\sigma) = \mathrm{const.}\equiv X_c.
\ee
One can see that this plane is invariant simply by observing that the canonical momentum corresponding to the off-plane motion is zero: $p'_X = \partial \mathcal L_\mathrm{eff} / \partial X_1 = 0.$ The corresponding NVE along the $X_1$-direction is trivial: $\delta X_1'' = 0$, yielding the conclusion that the system is integrable. We have seen that the system nevertheless exhibits an exponential growth of the in-plane variation with a positive Lyapunov exponent in the near-horizon region. By itself this is not surprising: a local instability can lead to a growing mode even in a trivially integrable system, the simplest example being the inverse chaotic oscillator \cite{Hashimoto:2020xfr}. This is similar to findings of \cite{Giataganas:2021ghs} where it is noted that horizons are really sources of instability in the bulk. Even integrable systems can display local instability in the vicinity of thermal horizons.

We need to make one thing clear. The integrability of the static open string Lagrangian (\ref{lag1}) that we have demonstrated in no way conflicts the established nonintegrability of string motion in black hole and D-brane backgrounds proved in \cite{StringsNonIntTseyt,StringsNonIntCherv}. The fact that a ring string in these backgrounds is nonintegrable, as found in the aforementioned references, is enough to prove the nonintegrability of string motion in these geometries in general, and this likely holds also for open strings with sufficiently complicated boundary conditions. On the other hand, the existence of special solutions and boundary conditions which are integrable (and therefore certainly nonchaotic) is perfectly possible also in a nonintegrable system.

\section{Open string in other backgrounds}\label{secother}

In order to further corroborate the universality of the result (\ref{2pit}), we will repeat exactly the same analysis for two more backgrounds: (1) a general hyperscaling-violating metric and (2) a black Dp brane. The former is quite relevant for many holographic purposes, the latter does not in general have a holography dual (as its asymptotic geometry is in general not AdS) but it does appear as a sector in various backgrounds (and of course the AdS throat of the D3 brane provides the simplest and most famous top-down AdS/CFT construction). This endeavour might look like mere stamp collecting but it has a purpose: to show that the result is not special to AdSS metric and also to show (from the Dp case) that by itself it has nothing to do with holography or AdS asymptotics -- it is all about thermal horizons.

\subsection{General hyperscaling-violating background}\label{subsechyper}

In our first example we closely follow the idea of \cite{Giataganas:2021ghs} and study bulk motion in a broad class of bulk geometries: hyperscaling-violating horizons at finite temperature, constructed in \cite{Goldstein:2009cv,Charmousis:2010zz,Gouteraux:2011ce,Gouteraux:2012yr} as gravity duals of effective field-theories with scaling and long-range entanglement, thought to be ubiquitous in quantum-many body systems. In \cite{Giataganas:2021ghs}, it is shown that the bulk geodesics, i.e. particle orbits also have the $2\pi T$ Lyapunov exponent in a large part of the parameter space though not everywhere; here we show that static strings/holographic heavy quarks \emph{always} yield $2\pi T$. The background metric reads
\be 
\label{hypermetric}ds^2=-r^{2\mathfrak{z}-\frac{2\theta}{D-2}}f(r)dt^2+\frac{1}{f(r)r^{2+\frac{2\theta}{D-2}}}dr^2+r^2d\vec{x}^2,\quad f(r)=1-\left(\frac{r_h}{r}\right)^{D-2+\mathfrak{z}-\theta},
\ee
and depends on two parameters, the Lifshitz exponent\footnote{We denote the Lifshitz exponent by $\mathfrak{z}$ rather than the usual $z$, as $z$ will be used for the radial coordinate $z=1/r$. Likewise $\zeta$ is taken by another coordinate to be used in Section \ref{secd1d5p}.} $\mathfrak{z}$ that measures the anisotropy of space versus time scaling (so that Lorentz-invariant backgrounds correspond to $\mathfrak{z}=1$) and the hyperscaling exponent $\theta$ which measures the deviation from the dimensional scalinig of free energy and roughly corresponds to long-range-entangled degrees of freedom. By definition, the temperature of the horizon at $r_h$ is found as:
\be
4\pi T=-\frac{g'_{tt}(r_h)}{\sqrt{g_{tt}(r_h)g_{rr}(r_h)}}=(D-2-\theta+\mathfrak{z})r_h^\mathfrak{z}.\label{temphyper}
\ee
We can easily redo the same analysis as for the AdSS configuration, keeping the same ansatz~(\ref{ansatzpolyak}) and the equations of motion analogous to (\ref{eomtxj}-\ref{vir22}). When everything is said and done, we obtain the near-horizon variational equation
\be 
\label{vareqhyper}\delta R''(\sigma)-(D-2-\theta+\mathfrak{z})^2r_h^{2\mathfrak{z}}\delta R(\sigma)=0,
\ee
which, according to (\ref{temphyper}), implies again $\lambda_L=2\pi T$ with the ansatz $\delta R\sim\exp(2\lambda_L\sigma)$. 

\subsection{Dp brane and related backgrounds}\label{subsecdbrane}

\subsubsection{Extremal black Dp brane}\label{subsubsecdbrane0}

Consider first the single Dp brane geometry in 10 spacetime dimensions. To the best of our knowledge no systematic work was done on string dynamics in brane backgrounds, except for the general proofs of nonintegrability in \cite{StringsNonIntTseyt,StringsNonIntCherv} and of course the near-brane limit of the D3 brane when the asymptotics becomes AdS. The metric reads
\begin{eqnarray}
ds^2 &=& \frac{\eta_{\mu\nu} dx^{\mu}dx^{\nu}}{f^2(r)} + f^2(r) \left(dr^2 + r^2 d \Omega_k\right),\quad \mu=0,\ldots p \\
f(r) &=& \left(1 + \frac{Q}{r^n} \right)^m, \quad n=7-p,k=8-p,m=\frac{1}{4}.\label{dbranemetric}
\end{eqnarray}
Here, $r$ is the radial coordinate, $x^\mu$ are the directions on the brane, while $d\Omega_k$ is the $k$-sphere with coordinates $\Phi_1,\ldots\Phi_k$. The string configuration we consider is completely analogous to the static open string studied previously in AdS black hole backgrounds:
\begin{eqnarray} \nonumber
    t&=&t(\tau), \quad X_1 = x_1, \quad \ldots \quad X_p=x_p, \\
    R&=&R(\sigma), \quad \Phi_1 = \phi_1,\quad \ldots\quad \Phi_k = \phi_k\label{dbranestringconfig}
\end{eqnarray}
Therefore, our string configuration is only nontrivial in the directions transverse to the brane, and reduced to a point on the $k$-sphere (i.e., in longitudinal directions). In Appendix \ref{secappa} we show that more general ans\"atze are possible; but for our purposes, Eq.~(\ref{dbranestringconfig}) is perfectly sufficient. For the time direction we can again choose $t(\tau) = \tau$. We have one nontrivial Virasoro constraint which, upon plugging into the equation of motion, yields:
\begin{equation}
R''+\frac{2f'(R)}{f(R)^5}-\frac{\left(R'\right)^2f'\left(R\right)\left(-f\left(R\right)^4+QR^{p-7}+1\right)}{f(R)^5}=0.
\label{dbraneeomr}
\end{equation}
We have some analytical control over the Eq.~(\ref{dbraneeomr}) and its variational equation for small $R$, i.e. near the brane, when we expand $R(\sigma)=\epsilon+\delta R(\sigma)$. The variational equation then reads
\be
\delta R''-\epsilon^{5-p}\sqrt{\frac{(7-p)(6-p)}{2Q}}\delta R=0,\label{dbranevar}
\ee
where $R\sim\epsilon$, i.e. the small parameter is the distance from the brane; since this limit means $\epsilon\to 0$, the bulk Lyapunov exponent vanishes. We will see the opposite situation with black Dp branes, in the presence of a thermal horizon. The solution (\ref{dbranevar}) is obviously only sensible for $p\leq 5$ but for $p>5$ the same conclusion is reached, only the expansion in $r$ large ($\epsilon$ small) is different.

\subsubsection{Non-extremal black Dp brane}\label{subsubsecdbranet}

We will now consider a non-extremal black brane, the finite-temperature generalization of the extremal solution at temperature $T$:\footnote{For concreteness we again assume $p\leq 5$ but, just like for the extremal brane, the generalization for $p>5$ is easy.}
\begin{eqnarray}
ds^2 &=&-h(r) \frac{dt^2}{f^2(r)} + \frac{d\vec x^2}{f(r)^2} + f^2(r) \left(\frac{dr^2}{h(r)} + r^2 d \Omega_k^2\right) \\
f(r)&=&\left(1+\frac{Q}{r^n}\right)^m,~h(r) = 1 - \left(\frac{r_{h}}{r}\right)^4, \quad n=7-p,k=8-p,m=\frac{1}{4}\label{dblackbranemetric}\\
\frac{1}{T} &=& \frac{4 \pi f(r_h)}{\sqrt{h'(r_h)(h(r)/f^2(r))'\big |_{r=r_h}}}=\frac{4\pi r_h\sqrt{1+Qr_h^{p-7}}}{p+1}. \label{dblackbraneh}
\end{eqnarray}
In the limit $r_h \rightarrow 0$ (equivalently, $T=0$) the black brane becomes the previously studied extremal black brane. The coordinates are the same as in the extremal solution (\ref{dbranemetric}). The $d\Omega_k$ sector is insensitive to temperature, which can be seen from the fact that its metric is independent of the redshift function $h$. The effect of the thermal horizon is thus seen solely in the equation of motion for $R$ (after plugging in the Virasoro constraint as usual):
\begin{equation}
R''-\left[\frac{f'(R)\left(-f\left(R\right)^4+QR^{p-7}+1\right)}{f(R)^5}+\frac{h'(R)}{2h(R)}\right]R'^2-\frac{h(R)\left(f\left(R\right)h'\left(R\right)-4h\left(R\right)f'\left(R\right)\right)}{2 f(R)^5}=0.\label{dblackbraneom}
\end{equation}
Following the same logic as before, we find the near-horizon variational equation:
\begin{equation}
\delta R''-\frac{(p+1)^2}{r_h^2+Qr_h^{p-5}}\delta R=0 .
\end{equation}
Looking for a solution of a form $\sim \exp(2 \lambda_L \sigma)$ and using Eq.~(\ref{dblackbraneh}), we find that $\lambda_L=2\pi T$. The ubiquitous $2\pi T$ is present even if the metric is asymptotically flat, as thermal horizons generate instability, whatever the faraway asymptotics. The holographic meaning of this instability is theory-dependent, and in general does not exist when there is no AdS region.

\subsubsection{From AdS${}_{p+2}\times\mathbb{S}^k$ throat to flat space}\label{subsubsecdbraneflat}

Following \cite{Gibbons:1993sv,Boonstra:1997dy,StringsNonIntTseyt}, we can consider a modification of the Dp brane geometry at zero or finite temperature to obtain a metric interpolating from an AdS${}_{p+2}\times\mathbb{S}^k$ throat (near-brane) to flat space in the far region; such solutions appear as solutions of supergravity and interpolate between different vacua. The expressions for the metric remain the same as before (Eq.~(\ref{dbranemetric}) at $T=0$ or Eq.~(\ref{dblackbranemetric}) at $T>0$) but the parameters are:
\be
m=\frac{1}{n},\quad p,n,k\textrm{ arbitrary}.
\ee
One can easily check that indeed for $Q\to 0$ or equivalently $r\to r_h$ (including the case $r_h=0$ for the extremal brane) we obtain AdS${}_{p+2}\times\mathbb{S}^k$ and for $Q\to\infty$ or equivalently $r\to\infty$ we get flat space (in fact, its product with the $k$-sphere). We proceed along the same lines as before, hence we only give the end results. The variational equation reads
\be 
\delta R''-\left(\frac{p+1}{r_h}\right)^2\left(1+Qr_h^{p-7}\right)^{\frac{4}{p-7}}\delta R=0,
\ee
which yields once again $\lambda_L=2\pi T$, computing the temperature by definition, similar as in Eq.~(\ref{dblackbraneh}). Now we can however obtain an analytic solution also in the far region, which interpolates to $\mathbb{R}\times\mathbb{R}^{p+1}\times\mathbb{S}^k$. Writing $R(\sigma)=1/\epsilon+\delta R(\sigma)$ with $\epsilon\to 0$, we get the variational equation
\be 
\delta R''+2(8-p)Q\epsilon^{9-p}\delta R=0,
\ee
yielding $\lambda_L=0$ as the coefficient of $\delta R$ is positive (hence the dynamics is oscillatory rather than exponentially growing), and in addition it drops to zero as we reach infinity.

All these examples strongly suggest that the $2\pi T$ exponent (with the same value, but different meaning from the MSS bound for CFT chaos) for the unstable saddle point in bulk motion is present if and only if the geometry has a static thermal horizon, holographic or not. In the next section we will see that the presence of rotation changes the outcome. This will lead us to the general conclusion: that the bulk near-horizon Lyapunov exponent is universal for a given symmetry class of the metric. For maximally symmetric horizons (isotropic and static) the result is always $2\pi T$, which is argued in detail also in \cite{fake:geometry}, on the basis of \cite{Lin:2019qwu}.

\subsubsection{Numerical solutions}\label{subsubsecdbranenum}

Although the equations we solve so far are quite elementary, it is always nice to have a numerical check too. Therefore, in this subsubsection we solve the equations of motion numerically; also, this is the only way to look at the string in whole space.\footnote{Remember we have no analytic control at intermediate distances, far from both the brane/horizon and the infinity/AdS boundary (depending on the geometry).} We will see explicitly that the system is integrable and yet that the variational equations have exponentially growing solutions.

\begin{figure}
    \centering
    \includegraphics[width=.9\linewidth]{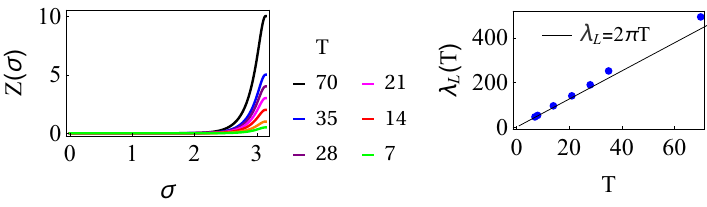}
    \caption{Radial profile $Z(\sigma)$ (A) and the numerical Lyapunov exponent $\lambda_L^{(n)}=\log\left(\delta Z\left(\sigma_0\right)/\epsilon\right)/2$ (B) for the static open string in thermal black brane background, for a range of temperatures $T$. We take $\sigma_0=0.2$ for the cutoff but values between $0.1$ and $0.5$ yield similar results. We compare the numerical result for the Lyapunov exponents to the analytic estimate (i.e., the chaos bound) and find reasonable agreement.}
    \label{figbbrane}
\end{figure}

We will look at the D3 black brane, The aim is thus to solve Eq.~(\ref{dblackbraneom}) and its variational equation for $p=3$.\footnote{The value $p=3$ is chosen as the D3 brane is particularly relevant for applications, having also the AdS throat, but the numerics works the same way for any $p$.} In order to do so, it is more convenient to use the coordinate $z=1/r$, so that the equation of motion for the new worldsheet field $Z(\sigma)$ becomes
\begin{equation} \label{mainEqZ}
    Z''-\frac{{Z'}^2}{Z}-\frac{h'(Z){Z'}^2}{2 h(Z)}-\frac{h^2(Z)Z^3}{f^4(Z)} \left(1-Z\left(\frac{2f'(Z)}{f(Z)}+\frac{h'(Z)}{2h(Z)}\right)\right) = 0.
\end{equation}
We impose Dirichlet conditions at the brane and Neumann conditions at the other end (open strings should be attached to branes but they can float freely in the asymptotically flat outer region). Once we have the solutions $Z(\sigma)$ we substitute the solution into the variational equation:
\begin{eqnarray}
\delta Z'' &-& \frac{\left( 2 h(Z)+Zh'(Z) \right)Z'}{Z h(Z)}\delta Z' +\frac{1}{2 Z^2}\Bigg[\frac{{Z'}^2 \left( 2 h^2(Z)+ Z^2 {h'}^2(Z)- h(Z) Z^2 h''(Z) \right)}{h^2(Z)} \nonumber\\
&-& \frac{20 v^2 h^2(Z)Z^6{f'}^2(Z)}{f^6(Z)}+ \frac{4 v^2 h(Z)Z^5 \left( 3 f'(Z) \left( 2 h(Z)+Zh'(Z) \right) + h(Z)Zf''(Z) \right)}{f^5(Z)} \nonumber\\
&& -\frac{v^2 Z^4 \left( 6 h^2(Z)+Z^2{h'}^2(Z)+h(Z)Z\left( 8h'(Z)+Zh''(Z) \right) \right)}{f^4(Z)}\Bigg] \delta Z = 0.
\end{eqnarray}
For $\delta Z$ the meaningful boundary condition is the fixed (and small) difference between the on-shell trajectory and its clone at the brane ($\delta Z(\sigma=0)=\epsilon$) and the Neumann condition at infinity (since the strings float freely so does the difference between to string profiles). The outcome is given in Fig.~\ref{figbbrane}. Along with the radial profiles of the string for different temperatures, we plot the near-horizon values of the numerically computed Lyapunov exponent $\lambda_L^{(n)}=\log\left(\delta Z\left(\sigma_0\right)/\epsilon\right)/2$, where $\sigma_0$ is some near-brane cutoff (we want the Lyapunov exponent near the brane thus $\sigma_0$ should cut off the far-from-brane part).\footnote{Of course, we have checked that the results do not strongly depend on $\sigma_0$.} The numerics is reasonably close to the analytic result, providing an additional confirmation.

\section{Open string in D1-D5-p black string background}\label{secd1d5p}

So far we have explored the bulk instability of open strings in black hole and black brane backgrounds and we have found the saturation of a fake MSS bound -- fake as it is unrelated to chaos (the configuration we consider is even integrable, though a more involved setup likely would not be such). Now we will interpret this finding and relate it to the thermal correlators and quasi-normal modes in a theory which is particularly interesting as we know something not only about the (super)gravity solution and the dual large-N field theory, but also about the microscopics: the D1-D5-p black string \cite{Greybody:1996,Strominger:1996sh,Maldacena:1997re,Seiberg:1999xz,MAGOO:1999,Lunin:2012gz}. This setup is celebrated for being the first black hole solution in string theory for which the entropy was computed by counting the microscopic degrees of freedom, obtaining for a horizon area $A$ the famous Bekenstein-Hawking result $S=A/4$ \cite{Strominger:1996sh}. Another famous result is the calculation of the greybody factor in \cite{Greybody:1996}, the logical macroscopic extension of the entropy calculation. To remind, the greybody factor is obtained in \cite{Greybody:1996} as the absorption cross section for a wavepacket in the black string background. In holographic setups, where the relevant near-horizon dynamics is dual to a two-dimensional CFT, the absorption cross section can be obtained from the imaginary part of the retarded Green's function.

Our idea here is twofold. First, we study the Lyapunov stability in D1-D5-p background -- this will yield some surprises as the geometry has a global rotation with angular velocity $\Omega$; so far we have only studied static geometries. Second, we will relate the Lyapunov exponent to the retarded propagator in dual field theory and pinpoint what it tells us about the meaning of bulk instability.

\subsection{Holography of the D1-D5-p system: a reminder}

The background describing the D1-D5-p black string reads
\begin{eqnarray}
    ds^2 &=& \frac{1}{{\sqrt{f(r)}}}\left(-dt^2+dx_5^2+\frac{r_0^2}{r^2} \left(\cosh\Sigma dt + \sinh\Sigma dx_5\right)^2\right)+\nonumber\\
    &&+ \sqrt{f(r)}\left(\frac{dr^2}{h(r)}+r^2\left(d\psi^2+\sin^2\psi\left(d\theta^2+\sin^2\theta d\phi^2\right)\right)\right),\label{d1d5metric}\\
   f(r)&=&\left(1+\frac{r_1^2}{r^2}\right)\left(1+\frac{r_5^2}{r^2}\right), \quad h(r) = 1 - \frac{r_0^2}{r^2}.\label{d1d5fh}
\end{eqnarray}
As usual, $t$ and $r$ are the time and radial coordinate respectively, $\psi$, $\theta$ and $\phi$ are the angles on the 3-sphere, and $x_i$ ($i=1,\ldots 5$) are the Cartesian coordinates in the plane. This is a classical solution of ten-dimensional type IIB supergravity compactified on $T^5 \cong T^4\times\mathbb{S}^1$ \cite{Greybody:1996,Kiritsis:2019npv}. It is charged under the Ramond-Ramond field of the corresponding theory; since D1 and D5 branes are magnetically dual to each other we get electric and magnetic charges that are related to the radii $r_1$ and $r_5$, respectively.\footnote{One can think of these charges also as representing the number of copies in the stack of D1 branes compactified on $\mathbb{S}^1$ along the $x_5$ direction, and in the stack of D5 branes wrapping the whole $T^4\times\mathbb{S}^1$ manifold.} There is also an additional charge associated to the p-momentum along D1-brane, i.e. the Kaluza-Klein (KK) mode on $\mathbb{S}^1$, related to factors of $r_0^2\cosh^2\Sigma$ in the metric (\ref{d1d5metric}-\ref{d1d5fh}). In the dilute gas regime these three charges satisfy \cite{Greybody:1996}:
\be
r_0, r_0 \sinh\Sigma \ll r_1, r_5.
\ee
The solution is anisotropic and rotating for $\Sigma \neq 0$ as we see from the presence of non-vanishing $tx_5$-component in the metric (\ref{d1d5metric}), implying that we now have the left and right temperature:
\be
T_L=\frac{r_0e^\Sigma}{2\pi r_1r_5},~~T_R=\frac{r_0e^{-\Sigma}}{2\pi r_1r_5}.\label{d1d5tltr}
\ee
We may further define the "average" (Hawking) temperature $T$ as $1/T=(1/T_L+1/T_R)/2$. This temperature and entropy are given by
\be
\frac{1}{T}=\frac{2\pi r_1r_5\cosh\Sigma}{r_0}, \quad S= \frac{2\pi^2r_1r_5r_0\cosh\Sigma}{4}.\label{d1d5ts}
\ee
We will need a few more features of this solution. The first is that in the extremal case ($T, S \propto r_0=0$), also known as the extremal D1-D5 bound state system, the near-horizon geometry becomes AdS${}_3 \times\mathbb{S}^3$. We can show this by performing the coordinate transformation $t\rightarrow t / \varepsilon L,~r \rightarrow \varepsilon L r,~ x_5 \rightarrow x_5 / \varepsilon L$ in the metric~(\ref{d1d5metric}), where $L^2 = r_1 r_5$: in the limit $\varepsilon \rightarrow 0$ we recover the AdS${}_3 \times\mathbb{S}^3$ geometry
\begin{equation} \label{ads3s3}
    ds_{\text{NHE}}^2 \approx \frac{r^2}{L^2} \left( - dt^2 + dx_5^2 \right) +  L^2 \frac{dr^2}{r^2} + L^2 d\Omega_3^2.
\end{equation}
On the other hand in the near-extremal case ($r_0 \rightarrow 0, \ {} \Sigma \rightarrow \infty$), the p-momentum survives and we still have the full D1-D5-p system, with a near-horizon geometry of the rotating Banados-Teitelboim-Zanelli (BTZ) black hole:
\begin{equation} \label{btzs3}
    ds^2_{\text{NHNE}} \approx  \frac{r^2}{L^2} \left( - dt^2 + dx_5^2 \right) + L^2  \frac{dr^2}{r^2-r_0^2} + \frac{r_0^2}{L^2} \left( \cosh \Sigma dt + \sinh \Sigma dx_5 \right)^2  + L^2 d\Omega_3^2.
\end{equation}
The procedure to derive this is the same as in the extremal case, except that now we also need to take $r_0 \rightarrow \varepsilon L r_0$. In order to translate the metric~(\ref{btzs3}) into the standard coordinates for BTZ black holes we have to perform an additional coordinate transformation:
\begin{equation}
\label{standCoord}
    r^2 = w^2 - w_-^2, \quad w_+ = r_0 \cosh \Sigma, \quad w_- = r_0 \sinh \Sigma.
\end{equation}
For convenience we will write down the metric of the rotating BTZ in these coordinates:
\begin{equation}
    ds^2_{\text{BTZ}} = - \frac{(w^2 - w_+^2)(w^2 - w_-^2)}{L^2 w^2}dt^2 + \frac{L^2 w^2 dw^2}{(w^2 - w_+^2)(w^2 - w_-^2)} + \frac{w^2}{L^2} \left( \frac{w_+ w_-}{w^2}dt + dx_5 \right)^2. \label{btzsc}
\end{equation}
The angular velocity is given by $\Omega = w_-/Lw_+ = \tanh \Sigma/L $.

From the above it is clear that the dual CFT lives in 1+1 spacetime dimension and encodes the physics of the AdS${}_3\times\mathbb{S}^3$ sector: in the holographic regime the large-$r$ flat asymptotics of the geometry (\ref{d1d5metric}) have to decouple. The classical gravity solution of course corresponds to the strongly coupled regime in CFT, however some direct comparisons were made (and count among the famous early tests of AdS/CFT) in the weakly-coupled regime where the CFT is approachable from field theory side \cite{Seiberg:1999xz,Lunin:2000yv,Lunin:2001pw,Lunin:2012gz} and some basic results from gravity side (like thermodynamics) still hold for reasons of continuity. In this regime the CFT has an orbifold point \cite{Lunin:2000yv,Lunin:2001pw} which acts as UV deformation, driving the theory toward a weakly coupled regime. This is seen in gravity as the deformation of the metric away from the near-brane region (AdS${}_3$ or BTZ). Therefore, we understand, at least to some extent, the UV physics and the meaning of the UV deformation of the theory.

\subsection{Radial fluctuations}\label{subsecd1d5rad}

We consider a static open string in D1-D5-p background stretching from interior to the boundary just like in previous examples. In particular, we postulate the following string configuration 
\begin{eqnarray}
    t(\tau,\sigma) &=& \tau, \quad R(\tau, \sigma) \equiv R(\sigma),\nonumber \\
    \Psi(\tau, \sigma ) &\equiv& \psi(\tau), \quad \Theta(\tau, \sigma) \equiv \pi/2, \quad \Phi(\tau, \sigma) \equiv \phi(\tau), \quad X_5(\tau, \sigma)\equiv  X_5(\sigma).\label{d1d5stringconfig}
\end{eqnarray}
The $\tau$-dependent degrees of freedom $\lbrace \Psi, \Phi \rbrace$ decouple. The remaining fields $R$ and $X_5$ also decouple from each other, since we can combine the equation of motion for $R$ with the nontrivial Virasoro constraint
\be \label{vir2}
    \frac{f(R) {R'}^2(\sigma)}{h(R)} + {X_5'}^2(\sigma) + \frac{r_0^2 \left( v^2 \cosh^2 \Sigma + \sinh^2 \Sigma {X_5'}^2(\sigma) \right)}{R^2(\sigma)} = 1
\ee
to obtain the following equation
\begin{eqnarray}
 &&2 f h^2 \left(-2r_0^2\cosh^2\Sigma+R^2\right)+ f' h^2 R \left(-r_0^2\cosh^2\Sigma + R^2 \right) + \nonumber\\
 &&+ f^2 R^2 \left(-\left(2h + Rh'\right){R'}^2 + 2hRR''\right) = 0.\label{lb1}
\end{eqnarray}
The effective Lagrangian for the coordinates $R$ and $X_5$ takes the form
\be \label{lagrd1d5p}
\mathcal L = \frac{1}{\sqrt{f(R)}}\left[\left(-1 + \frac{r_0^2 \cosh^2 \Sigma}{R^2} \right) - \frac{f(R) {R'}^2}{h(R)} - \left( 1 + \frac{r_0^2 \sinh^2 \Sigma}{R^2} \right) {X_5'}^2 \right],
\ee
and reproduces Eq.~(\ref{lb1}) when combined with the Virasoro constraint (\ref{vir2}).

We will assume that we are in the dilute gas regime, like in \cite{Greybody:1996}, defined by $r_0, r_0 \sinh\Sigma \ll r_1, r_5$. This boils down to the condition $T\ll 1/r_1,1/r_5$. As usual, we can solve the equation analytically in two distinct regions: (1) near-horizon region $r \sim r_0,r_0 \sinh\Sigma \ll r_1, r_5$ and (2) far region $ r_0,r_0 \sinh\Sigma \ll r \sim  r_1, r_5$.

Expectedly, the system described by the Lagrangian (\ref{lagrd1d5p}) is integrable. Applying the NVE methods, we can choose the invariant plane to be $\{ t = \tau, R = r_0, \Psi =0, \Theta = \pi/2, \Phi =0, X_5 = \mathrm{const.} \}$. Since $X_5$ is a cyclic coordinate in (\ref{lagrd1d5p}), its conjugate momentum is constant: $p'_{X_5} = \partial \mathcal{L}/\partial X_5 = 0$. Therefore, the $R-X_5$ plane is indeed invariant under the evolution of the system (along $\sigma$). The normal variational equation therefore corresponds to the variations in the $X_5$-direction, yielding $\delta X_5'' = 0$. Just like the open string dynamics on the Schwarzschild horizon, this system is integrable.\footnote{Again, a more general open string setup would not necessarily be integrable -- one has to perform the Kovacic analysis for every boundary condition (i.e. effective Lagrangian) separately.} In both cases, the extra integrals of motion are simply the transverse momenta.

We solve the variational equation of the radial coordinate in (\ref{lb1}) obtained by perturbing the horizon solution $R(\sigma)=r_0$ as $R \sim r_0 + \delta R(\sigma)$:
\bea
&& \delta R''-\frac{2\mathfrak{a^4}}{r_0^6 f^2(r_0)} \delta R =0,\label{d1d5horeom1}\\
&& \mathfrak{a^4} \equiv r_0^2(r_1^2+r_5^2)+2 r_1^2 r_5^2 + r_0^2\left(2 r_0^2 + r_1^2 + r_5^2\right)\cosh 2\Sigma\label{d1d5horeom2}
\eea
Plugging in the expression for $f$, we find the exponent of the asymptotic growth of the solution, determined as before by $\sim e^{2 \lambda_L\sigma}$:
\begin{equation}
\lambda_L=\frac{r_0}{r_1r_5}\sqrt{1+\frac{r_0^2(r_1^2+r_5^2)}{2r_1^2r_5^2}\cosh(2\Sigma)}.\label{d1d5ple}
\end{equation}
The above expression\footnote{One may worry whether this expression remains finite in the near-extremal limit where we take $\Sigma \rightarrow \infty$. We should pay attention to the fact that there is a factor of $r_0$ hiding inside the temperature $T$. In the near-extremal limit we also take a limit $r_0\rightarrow 0$, while keeping $r_0 \cosh \Sigma$ fixed.} can be written in terms of the temperatures (\ref{d1d5tltr}-\ref{d1d5ts}) as:
\be 
\lambda_L=2\pi T\cosh\Sigma\sqrt{1+\pi^2\left(r_1^2+r_2^2\right)\left(T_L^2+T_R^2\right)}=2\pi T\cosh\Sigma\left(1+\frac{\pi^2}{2}\left(r_1^2+r_5^2\right)\left(T_L^2+T_R^2\right)+\ldots\right).\label{d1d5ple1}
\ee
The second equality (expansion in $r_0^2/\sqrt{r_1^2+r_5^2}$) is the dilute-gas approximation. Importantly, the Lyapunov exponent \emph{does not repeat the universal $2\pi T$ ("fake MSS") result}. It depends on $r_1$ and $r_5$ in addition to $T$, and its temperature dependence is a nonlinear function. But the leading term in the expansion has a simple form:
\be
\lambda_L^{(0)}\approx\frac{r_0}{r_1r_5}=2\pi T\cosh\Sigma.\label{d1d5ple0}
\ee
Therefore, \emph{the Lyapunov exponent in the dilute-gas regime "comes close" to the static value but differs by a factor of $\cosh\Sigma$ which equals unity when $\Sigma = 0$, i.e. when there is no rotation}. In absence of rotation we return to the $2\pi T$ exponent in the dilute-gas regime.

We can translate our result into the standard variables for rotating BTZ solutions. Since in standard coordinates for BTZ black holes we have $\Omega = w_-/L w_+$, using Eq.~(\ref{standCoord}) it follows that $L \Omega = \tanh  \Sigma.$ Therefore, using the identity $\cosh \Sigma = 1/\sqrt{1 - \tanh^2 \Sigma}$, we get
\begin{equation} \label{lambdad1d5p}
    \lambda_L^{(0)} = \frac{2 \pi T }{\sqrt{1- L^2 \Omega^2}}, \quad L \Omega \in [0,1).
\end{equation}
We could express this result in terms of the left and right temperature $T_{L,R}$, making use of the relation $2/T=1/T_L+1/T_R ~ \Rightarrow ~ T = 2 T_L T_R / (T_L+T_R)$. However, we do not get a particularly simple or more intuitive form than (\ref{lambdad1d5p}), which in fact nicely shows how a nonzero rotation rate $\Omega$ deforms us away from the universal $2\pi T$ scaling.

This result should be compared to the ones found in \cite{Jahnke:2019gxr,Banerjee:2019vff}, where chaos in dual CFT was studied by calculating the OTOC correlators of rotating BTZ black holes for a scalar field and for a probe string respectively. The calculation done in \cite{Jahnke:2019gxr} obtains two different Lyapunov exponents $\lambda_L^{\pm} = 2 \pi T / (1 \mp L \Omega)$ in the presence of rotation, one of which is above the MSS bound and the other one bellow it, presuming that $\Omega \neq 0$. Our result~(\ref{lambdad1d5p}) turns out to be exactly equal to the geometric mean of $\{\lambda_L^+, \lambda_L^-\}$, implying that $\lambda_L^-<\lambda_L<\lambda_L^+$. Both results show that when an additional scale (angular velocity) appears there is no single "degenerate" exponent anymore, but different response functions receive different corrections.

We note in passing that our near-horizon analysis yields a single Lyapunov exponent, rather than a Lyapunov spectrum with two (in general different) exponents as one would expect in this background (and as \cite{Jahnke:2019gxr} finds in the rotating BTZ case) --  rotation breaks isotropy so the two directions normal to the invariant plane should be inequivalent. The reason that we nevertheless only see a single exponent could be that the quanta of p-momentum in D1-D5-p are only left-moving, thus we only see the Lyapunov exponent associated with the temperature of the left-moving modes.

The opposite limit when $r \gg r_0, r_0 \cosh \Sigma$ and $r \gg r_1, r_5$, is treated in the same way as the asymptotically flat limit of the interpolating geometry in subsubsection \ref{subsubsecdbraneflat}:  we have noted that we can think of the six-dimensional black string as an interpolation between AdS${}_3 \times\mathbb{S}^3$ and Minkowski spacetime. The far region corresponds to the latter, hence it must have zero Lyapunov exponent.

Unlike the examples from previous sections, we have now found some unexpected aspects of the bulk Lyapunov exponent: dependence on the rotation rate and the complex temperature dependence away from the dilute-gas limit.\footnote{There is of course no rigorous reason for the bulk exponent to obey the MSS bound in all cases as it is a different object, and comparing to \cite{Jahnke:2019gxr} we have seen that even though the OTOC exponent also changes in the presence of rotation the values are different. But still, the fact that (\ref{d1d5ple1}-\ref{d1d5ple0}) differ from the results like \cite{Jahnke:2019gxr} suggests there is a difference in underlying physics.} Now we will relate it to the retarded propagator of a CFT quasiparticle interacting with the thermal ensemble,\footnote{In the context of D1-D5 CFT it does not make much sense to talk about quarks as the symmetries of the theory differ from $\mathcal{N}=4$ SYM.} the natural dual object to consider.

\subsection{Transverse fluctuations}\label{subsecd1d5trans}

Now we want to study the retarded Green's function for transverse fluctuations, which describes thermal motion of quasiparticles in D1-D5 field theory, the more intuitive object to consider compared to the radial fluctuation. In this case, it is convenient to switch to the Nambu-Goto formalism and work in the static gauge. A similar calculation was already done in a slightly different setup \cite{BenerjeeFiniteDensity}, where the authors study the bulk dynamics of a fundamental string in an extremal and near-extremal Reissner-Nordstr\"{o}m (RN) black hole background. Of course, the D1-D5 background will give very different physics.

The ansatz is now:
\begin{eqnarray}
    t(\tau,\sigma) &=& \tau \equiv t, \quad R(\tau, \sigma) = \sigma \equiv r,\nonumber \\
    \Psi(\tau, \sigma ) &\equiv& \psi(t), \quad \Theta(\tau, \sigma) \equiv \pi/2, \quad \Phi(\tau, \sigma) \equiv \phi(t), \quad X_5(\tau, \sigma)\equiv  X(t,r).\label{d1d5stringconfig2}
\end{eqnarray}
We expand $X(t,r)$ in Fourier modes
\be
X(t,r) = \int \frac{d \omega}{2 \pi}e^{- i \omega t}X_{\omega}(r),
\ee
and the relevant equation of motion obtained by varying the Nambu-Goto action reads
\begin{eqnarray} \nonumber
    && X_{\omega}''(r) + \left( \frac{r_0^2 \left( -3 r^2 +r_0^2+ \left(r^2- r_0^2 \right)\cosh^2\left( 2 \Sigma \right) \right)}{r \left( r^2-r_0^2 \right)\left( - 2 r^2 +r_0^2 +r_0^2 \cosh^2 \left(2 \Sigma \right) \right)}  - \frac{f'(r)}{f(r)} + \frac{h'(r)}{2h(r)}\right) X_{\omega}'(r)  \\
    && - \frac{2 r^2 \omega^2 f(r)}{\left( -2 r^2+r_0^2 + r_0^2 \cosh^2\left( 2 \Sigma \right) \right) h(r)}X_{\omega}(r) = 0.       \label{eomxxfluct}
\end{eqnarray}
In the special case when there is no rotation ($\Sigma = 0$) this equation simplifies to
\begin{eqnarray}
    && X_{\omega}''(r) + \left( \frac{r_0^2}{r^3 h(r)}   - \frac{f'(r)}{f(r)} + \frac{h'(r)}{2 h(r)} \right) X_{\omega}'(r) + \frac{\omega^2 f(r)}{ h^2(r)}X_{\omega}(r) = 0.       \label{eomxxfluctnonrot}
\end{eqnarray}
We will first solve the special case with no rotation; it will give us some useful prior intuition for the general case. Since the problem can be divided into two regions, we will again employ the matching procedure in order to gain some analytic control of the equation.

\subsubsection{Static AdS: extremal case}\label{subsubsecd1d5norotext}

\begin{figure}
    \centering
    \includegraphics[scale=0.75]{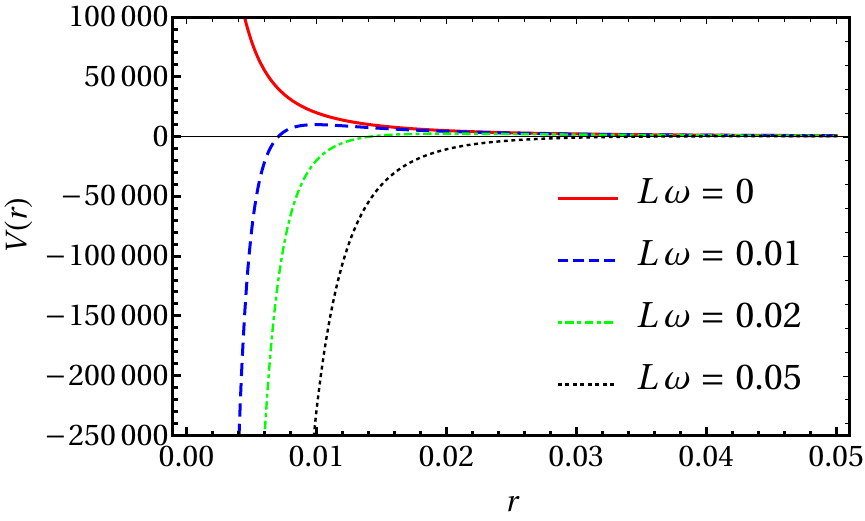}
    \caption{The effective Schr\"{o}dinger potential (\ref{veffext}) for the extremal D1-D5 geometry with $L=1$, for four values of the frequency $\omega$. The static case $\omega=0$ (red full line) is qualitatively different because the potential is strongly repulsive: there is no absorption because plan waves coming from infinity are reflected away. For $\omega>0$ (blue, green, black dotted lines) the potential is strongly attractive, diverging as $1/r^4$ at the origin $r=0$. This again implies zero absorption cross section as there are no solutions behaving as plane waves at infinity.}
    \label{schrdextr}
\end{figure}

Consider first the near-horizon region of the extremal black string, i.e. at temperature $T=0$ on the field theory side. In this case the IR geometry is given by Eqs.~(\ref{ads3s3}). The relevant equation of motion for string fluctuations along the $x_5$-direction in this regime is
\be
X''_\omega(r)+\frac{4}{r}X'_\omega(r)+\left(\frac{L^2\omega}{r^2}\right)^2X_\omega(r)=0, \label{eqextr}
\ee
with general solutions of the form
\be
X_{\omega}(r) = \mathcal{A} \left( 1 - \frac{i L^2 \omega}{r} \right)e^{\frac{i L^2 \omega}{r}} +  \mathcal{B}\frac{1}{2 L^4 \omega^2 r} \left(1  - \frac{i r }{L^2 \omega} \right)e^{-\frac{i L^2 \omega}{r}}.
\ee
Imposing the infalling boundary condition (appropriate for the retarded propagator) at the horizon requires $\mathcal{B} = 0$. Expanding this solution in the matching region $r_0 \ll r \ll L$, we get
\be
    X_{\omega}(r) = \mathcal{A} \left( 1 + \frac{\left(L^2 \omega\right)^2}{r^2} \right).
\ee
From this we can calculate the retarded Green's function at $T=0$ in the IR region $r_0/r,\omega L\ll 1\ll L/r$:
\be
    \mathcal G_R^{(T=0)}  = L^4 \omega^2 \Rightarrow \Im \mathcal G_R^{(T=0)} = 0.
\ee
Therefore, we get a vanishing absorption cross-section in the presence of the horizon, i.e. $\Im\mathcal G_R^{(T=0)} = 0$. So we need to understand why the extremal horizon does not absorb anything even though it is a horizon (with finite area and finite greybody factor). From the bulk viewpoint, one way to see the reason is to rewrite the fluctuation equation (\ref{eqextr}) in the Schr\"odinger form:
\be
\partial_r^2\widetilde X_\omega(r)-V_\mathrm{eff}(r)\widetilde X_\omega(r)=0,\quad V_{\text{eff}}(r) = \frac{ 2 r^2 - L^4 \omega^2 }{r^4}.\label{veffext}
\ee
The effective potential is shown in Fig.~\ref{schrdextr} for various values of $\omega$. For $\omega = 0$ a zero imaginary part could be expected -- for $\omega=0$ the effective potential is positive and (quadratically) divergent at the horizon, thus there is no absorption, i.e. all incoming waves are reflected backward.

The nonstatic case $\omega\neq 0$ is qualitatively different -- it has the expected negative divergence (infinite well) at the extremal horizon $r\to 0$, as we see from Eq.~(\ref{veffext}) and Fig.~\ref{schrdextr}. However, it is known that the scattering problem for attractive central potentials diverging as $1/r^s$ for $s>2$ is not well-defined \cite{Landau:1991wop}: such potentials always lead to a wave "falling toward the center" and the solution to the Schr\"odinger equation in this case is always localized around zero -- there is no absorption because the infalling plane wave at infinity is not a consistent boundary condition. We will see in the following section that we can infer this result for the retarded Green's function in the extremal case by considering the limit $\omega \ll T$ of the thermal correlator obtained in a near-extremal case.

\subsubsection{Static BTZ: near-extremal case}\label{subsubsecd1d5norotnearext}

\begin{figure}
    \centering
    \includegraphics[scale=0.75]{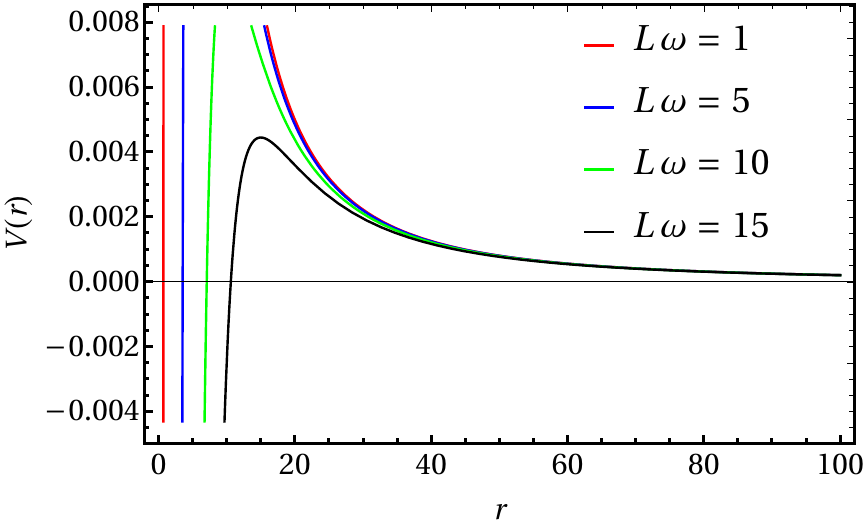}
    \caption{The effective Schr\"{o}dinger potential (\ref{btzveff}) for a near-extremal D1-D5-p system with the parameters $r_0=0.1,~ L=1$, for $L\omega=1,5,10,15$ (red, blue, green, black). Already from the expression in Eq.~(\ref{btzveff}) it is obvious that in the near-extremal case nothing special happens in the static limit $\omega=0$. For all frequencies, the potential has the form typical of near-horizon effective potentials \cite{MAGOO:1999,Kiritsis:2019npv}, where a high but finite potential barrier is followed by the infinite well at the horizon $r=r_0$.}
    \label{schrd}
\end{figure}

At low but finite temperatures or equivalently in the near-extremal case we would be interested in the dynamics of the open string in the metric given by Eq.~(\ref{btzs3}). Therefore, we look for the solution of open string equations in BTZ$\times\mathbb{S}^3$ geometry but (in this subsection) still with no rotation ($\Sigma=0$). The relevant equation can be written in a compact form reminiscent of the relativistic wave equation in curved background:
\be
\frac{h(r)}{r^4} \frac{d}{dr}\left( h(r) r^4 \frac{dX_{\omega}(r)}{dr}  \right) + \frac{L^4 \omega^2}{r^4} X_{\omega}(r) = 0.
\label{btzeqx}
\ee
It is again instructive to look at the Schr\"{o}dinger form of the equation, obtained by plugging in $X_{\omega}(r) = h^{-1/2}(r) r^{-2} \Psi(r)$ into Eq.~(\ref{btzeqx}):
\be
\left( \frac{d^2}{dr^2} - V_\mathrm{eff}(r) \right) \Psi(r) = 0,\quad V_\mathrm{eff}(r)=\frac{2 r^2 -3 r_0^2 - L^4 \omega^2}{\left( r^2 - r_0^2 \right)^2}.\label{btzveff}
\ee
The second term inside the brackets is the effective Schr\"{o}dinger potential, plotted in Fig.~\ref{schrd}.

Proceeding further toward the analytic solution to Eq.~(\ref{btzeqx}) it is convenient to transform the radial variable as 
\be
\zeta\equiv\frac{r_0^2}{r^2}.\label{zetacoord}
\ee
In order to reduce Eq.~(\ref{btzeqx}) to a hypergeometric differential equation,\footnote{Since Eq. (\ref{btzeqx}) has three regular singular points at $r=0,~r_0$ and $\infty$, we can be sure that it can be written in the form of the hypergeometric differential equation.} we will make a further coordinate transformation $\zeta \mapsto 1-\xi$. The equation now reads
\be 
X_{\omega}''(\xi)-\frac{1}{2\xi}\frac{2-\xi}{1-\xi}X_{\omega}'(\xi)+\frac{1}{\xi^2(1-\xi)}\left(\frac{L^2 \omega}{2 r_0}\right)^2 X_{\omega}(\xi)=0.
\label{btzeqxh}
\ee
We can solve this equation at the horizon $\xi=0$, by making the substitution $y = - \log\xi$ in Eq.~(\ref{btzeqxh}). The solution at the horizon takes the form $X_{\omega}\sim e^{\pm i \alpha y}= \xi^{\pm i \alpha}$, with $\alpha = L^2 \omega / 2 r_0$. The boundary condition at the horizon requires the outgoing modes to vanish, yielding
\be
X_{\omega}(\xi) = \widetilde{\mathcal{A}}~ \xi^{-i \alpha}. \label{bndc}
\ee
In order to get the full near-horizon solution, we plug the ansatz $X_{\omega}(\xi) = \xi^{- i \alpha}F(\xi)$ into Eq.~(\ref{btzeqxh}), yielding
\be
    \xi(1-\xi) \frac{d^2F}{d\xi^2} + \left[ 1 - 2 i\alpha  - \left(1 - i\alpha - \frac{1}{2}  -i \alpha \right)\xi \right] \frac{dF}{d\xi} -(- i\alpha)\left(- \frac{1}{2} - i\alpha \right)  F(\xi) = 0.\label{d1d5nearhyper}
\ee
We recognize Eq.~(\ref{d1d5nearhyper}) as the hypergeometric equation with parameters
\be
a = - i\alpha,\quad b= -\frac{1}{2}-i \alpha, \quad c = 1 - 2 i \alpha.
\ee
The corresponding regular solution reads
\be
    F(\xi) = \widetilde{\mathcal {A}} \ {}_2F_1 \left(a,b,c; \xi \right) + \widetilde{\mathcal {B}} ~ \xi^{2 i \alpha} \ {}_2F_1 \left( a + 1 -c,b+1-c,2-c ; \xi \right).
\ee
We impose the infalling boundary condition (\ref{bndc}) at the horizon, implying that $\widetilde{\mathcal {B}} = 0$, thus the near-horizon solution becomes
\be
    X_{\omega}(\xi) = \widetilde{\mathcal {A}}~ \xi^{- i \alpha} \ {}_2F_1 \left(a,b,c; \xi \right). \label{d1d5pfinsol}
\ee
We can now use the following identity to express solution (\ref{d1d5pfinsol}) in terms of functions depending on $\zeta$, instead of $\xi = 1-\zeta$:
\begin{eqnarray} \nonumber
    \ {}_2F_1 \left(a,b,c; \xi \right)& = &\frac{\Gamma (c) \Gamma(c-a-b)}{\Gamma(c-a)\Gamma(c-b)} \ {}_2F_1 \left(a,b,1 + a+b -c; \zeta \right)+ \\
    &&+ \frac{\Gamma(c)\Gamma(a+b-c)}{\Gamma(a)\Gamma(b)}\zeta^{c-a-b} \ {}_2F_1 \left(c-a,c-b,1+c-a-b; \zeta \right).~~~ \label{idty}
\end{eqnarray}
The matching region $r_0 \ll r$ corresponds to $\zeta \ll 1$, so we expand Eq.~(\ref{idty}) in small $\zeta$:
\be \label{2F1exp}
    \ {}_2F_1 \left(a,b,c; \xi \right) \approx \frac{\Gamma (c) \Gamma(c-a-b)}{\Gamma(c-a)\Gamma(c-b)} + \frac{\Gamma(c)\Gamma(a+b-c)}{\Gamma(a)\Gamma(b)}\zeta^{c-a-b}, 
\ee
We observe that the full solution in the matching region is of the form
\be
    X_{\omega}(r) \propto \widetilde{\mathcal S}~r^{-d + \Delta} + \widetilde{\mathcal F}~r^{-\Delta}, \quad d=\Delta = 3,
\ee
which allows us to read off the retarded Green's function as the ratio $\widetilde{\mathcal F}/\widetilde{\mathcal S}$:
\be
    \mathcal{G}_R^{(T)}(\omega) \propto \frac{\Gamma(c-a)\Gamma(c-b)}{\Gamma(a)\Gamma(b)} = \frac{\Gamma\left(1-i \frac{\omega}{2\lambda_L}\right)\Gamma\left(\frac{3}{2}-i \frac{\omega}{2\lambda_L}\right)}{\Gamma\left( - i \frac{\omega}{2\lambda_L}\right)\Gamma\left(-\frac{1}{2}- i \frac{\omega}{2\lambda_L}\right)}. \label{2ptcorr}
\ee
We can take the imaginary part of (\ref{2ptcorr}) to get the absorption cross-section:
\be
    \sigma_{\text{abs}} = \Im \mathcal{G}_R^{(T)}(\omega) \propto \frac{\alpha}{4} + \alpha^3, \quad \alpha = \frac{\omega}{2\lambda_L}, \quad \lambda_L = 2 \pi T.\label{2ptcorrim}
\ee
This is the central result of our calculation -- the IR propagator and the absorption cross section for a heavy quasiparticle excitation. Let us think what this result means:
\begin{enumerate}
    \item The only energy scale in the Green's function is the temperature. This is in line with the problem of drift of a heavy quark through neutral $\mathcal{N}=4$ super-Yang-Mills (SYM) plasma, dual to a dragging string in AdSS background \cite{HerzogDragForce,GubserDragForce,Gubser:2006nz} and many subsequent works in the same setup \cite{Giataganas:2013hwa,Giataganas:2018rbq,Casalderrey-Solana:2007ahi,Casalderrey-Solana:2009ifi,Atmaja:2010uu}: in this case the quasiparticle does not see the charges of the D1-D5-p system.\footnote{ At least, this is the case in our current setup with no drift; it would be interesting to check if this conclusion remains in force in presence of drift.} This can be ascribed to the fact that the additional charges of the D1-D5-p system are global and the quasiparticle is neutral with respect to them.
    \item The form of the propagator (Eq.~\ref{2ptcorr}) could be expected from the BTZ asymptotics of the near-extremal geometry \cite{Blake:2019otz}, as it has the form of conformal quantum mechanics, i.e. 0+1-dimensional CFT \cite{Faulkner:2009wj} (we know that in the near-horizon region of the BTZ geometry the transverse spatial coordinate decouples and the geometry becomes AdS${}_2\times\mathbb{S}$, so that AdS${}_2$ gives the 0+1-dimensional CFT).
    \item The imaginary part behaving as $\sim\omega+\omega^3$ suggests that in addition to the usual drag force $f\propto\dot{x}$ we also have a third-order term $\tilde{f}\propto dx^3/dt^3$. This is in fact expected -- all odd-power terms\footnote{Even-power terms (like $\dot{x}^2$) are not expected as their sign is independent of the sign of velocity, i.e. a proper drag force (opposing the motion) would have to look like $-\dot{x}^2\mathrm{sgn}\dot{x}$ but that implies the breaking of some discrete symmetry which we do not have.} in velocity are allowed symmetry-wise and the leading-order holographic Green function already captures the first two terms.
    \item This result could not be reproduced neither from the static limit ($\omega=0$) nor from the extremal limit $T=0$ -- these two limits are singular, which is expected for the static limit but somewhat strange for the extremal limit.
\end{enumerate}

As a sanity check we consider the high-frequency limit $\omega \gg T$ where one should recover the result for the extremal case.\footnote{In this limit we consider wavelengths well below $T^{-1}$ ($\omega^{-1} \ll T^{-1}$) that are insensitive to thermal fluctuations and thus resemble the behavior for the extremal background geometry.} In this limit we can compare our calculation to the pure CFT result for the two-point correlation function. We consider a two-point correlation function in a $2+1$-dimensional CFT for an operator with scaling dimension $\Delta$: this behaves as $\langle \mathcal{O}_{\Delta}(t) \mathcal{O}_{\Delta}(0) \rangle \sim |t|^{-2\Delta}$, i.e. $\sim \omega^{2 \Delta - d}$, where $d$ is the spacetime dimensionality. For an operator with a scaling dimension $\Delta = 3$ living in $d=3$ spacetime dimensions, one should indeed expect $\sim\omega^3$ power-law behavior of the thermal correlator in the high-frequency limit.

\subsubsection{Rotating BTZ: near-extremal case}\label{subsubsecd1d5rotnearext}

Now we study the rotating BTZ black hole. Compared to the static case, it is considerably more difficult for calculations. Conceptually, it is also distinct for having different left and right temperature. Now we cannot write an analytic solution in the whole throat, all the way to the AdS boundary (i.e., the throat of the D1-D5-p geometry, before the far-region flat asymptotics kick in), akin to Eq.~(\ref{d1d5pfinsol}). Instead, we can only treat the near-horizon limit, when $r\to r_h$ or equivalently $\zeta\to 1$ (for $\zeta$ as defined in Eq.~(\ref{zetacoord})).

The equation of motion for the transverse fluctuations (\ref{eomxxfluct}) can be slightly rewritten as
\be
    \left(1 - \frac{r_0^2}{r^2} \right) \left(1 - \frac{r_0^2+r_n^2}{r^2} \right) X_{\omega}''(r) + \frac{4}{r}\left[ 1 - \frac{3}{2} \left( 1 - \frac{r_0^2}{3r^2} \right) - \frac{5}{4}\left( 1- \frac{2}{5} \frac{r_0^2}{r^2} \right) \right] X_{\omega}(r) + \frac{L^4 \omega^2}{r^4}X_{\omega}(r) = 0.\label{d1d5rot}
\ee
Now we perform the same change of variables as before, $y=-\log\xi=-\log\left(1-r_0^2/r^2\right)$, so that $y\rightarrow\infty$ at the horizon. Expanding the coefficients to order $\xi^0$ for $\xi$ small or equivalently to order $1=(e^{-y})^0$ for $y$ large, we get the near-horizon limit of (\ref{d1d5rot}):
\be
    \left(1 + \frac{2r_n^2}{r_0^2} - e^y \frac{r_n^2}{r_0^2}\right)X_{\omega}''(r)  + e^y \frac{r_n^2}{2r_0^2}X_{\omega}'(r) + \frac{L^4\omega}{4 r_0^2} X_{\omega}(r) =0.
\ee
The solution that satisfies the infalling boundary condition at the horizon is
\be
    X_{\omega}(y) \sim e^{i \delta y} \ {}_2F_1 \left(i \delta, -\frac{1}{2} + i \delta, 1+ 2 i \delta; \frac{e^y r_n^2}{r_0^2 + 2 r_n^2}\right), \quad \delta = \frac{L^2 \omega}{2 \sqrt{r_0^2 + 2 r_n^2}}.
\ee
Another way to understand the fact that in the presence of rotation the solution of this type cannot exist without imposing some approximations is that hypergeometric functions are the representations of $\mathrm{SL}(2,\mathbb R)$ which is broken by rotation.

In order to obtain the IR propagator (now we cannot obtain analytically the propagator for general $\omega$ values), we expand the above solution in the region far from the horizon, for $r \gg r_0$, that is for $y$ small. Using the identity (\ref{2F1exp}) this yields
\be
    \ {}_2F_1 \left(i \delta, -\frac{1}{2} + i \delta, 1+ 2 i \delta; \frac{e^y r_n^2}{r_0^2 + 2 r_n^2}\right) \sim \frac{\Gamma\left(\frac{3}{2}\right)\Gamma(1+ 2 i \delta)}{\Gamma(1+ i \delta)\Gamma\left(\frac{3}{2}+ i \delta\right)} + \frac{\Gamma\left(1+2 i \delta\right)\Gamma\left(-\frac{3}{2}\right)}{\Gamma(i \delta)\Gamma\left(-\frac{1}{2}+ i \delta\right)}\left(\frac{r_0}{r}\right)^{3}.
\ee
Identifying the leading term and subleading term as the source and response respectively we compute the retarded Green's function in the rotating case:
\be
    \mathcal G_R^{(T)}(\omega) \sim \frac{\Gamma(1- i \delta)\Gamma\left(\frac{3}{2}- i \delta\right)}{\Gamma(-i \delta)\Gamma\left(-\frac{1}{2}- i \delta\right)}
\ee
It has the same form as the one that we have obtained in previous section in the absence of rotation (\ref{2ptcorr}), but with a different $\lambda_L$ scale compared to Eq.~(\ref{d1d5ple0}):
\be
    \lambda_L^{\text{(mod)}} = 2 \pi  T \cosh \Sigma \sqrt{1 + \frac{2r_n^2}{r_0^2}} = \lambda_L^{(0)} \sqrt{\cosh(2 \Sigma)}, \quad \delta = \frac{\omega}{2 \lambda_L^{\text{(mod)}}},
\ee
We can interpret this as a modification of energy in the presence of a rotating horizon, i.e. the Lense-Thirring effect.

\subsection{Quasinormal modes and their decay scale}
\label{QNMsSec}
So far, we have found that breaking a global symmetry in IR (introducing rotation) in general influences the bulk instability scale differently from the way it influences the exponent of the field-theory OTOC. We have likewise seen that UV deformations (full brane geometry deforming the AdS throat) also change the instability scale away from $2\pi T$. It remains unclear however what exactly the bulk scale is from the CFT viewpoint, and to understand this we will relate the bulk instability exponent to the quasinormal mode frequencies. For simplicity we will consider the non-rotating case but all the results that we obtain still hold also in the presence of rotation with $\lambda_L \rightarrow \lambda_L^{\text{(mod)}}$.

Let us first remember that the poles in the retarded Green's function are related to transport properties of a thermal field theory. On the gravity side, they correspond to a spectrum of quasinormal modes \cite{MAGOO:1999,Kiritsis:2019npv}. More specifically, the relaxation times in field theory are given by the imaginary part of the QNM spectrum in the bulk \cite{Cardoso:2001hn, Berti:2009kk}. Since our retarded Green's function (\ref{2ptcorr}) is singular at an infinite number of points in the complex plane, due to the presence of the gamma functions in the numerator, we can extract the whole QNM spectrum from it. Singular points are given by $c-a = -n$ or $c-b = -n$, for $n \in \mathbb Z^+$ (a set of non-negative integers), thus $\omega_n = -2i (n+1) \lambda_L$ or $\omega_n = - 2i (n+3/2) \lambda_L.$ The union of the two sets yields the following spectrum:
\be
\omega_{\mathfrak n} = -2i (\mathfrak n + 1) \lambda_L, \quad \mathfrak n = 0, \frac{1}{2},1,\frac{3}{2}, \cdots \label{spectrum0}
\ee
or equivalently
\be
\omega_{\mathfrak m} = -i(\mathfrak m + 1) \lambda_L, \quad \mathfrak m = 1,2,3, \cdots. \label{spectrum}
\ee
We write the solution in these two obviously equivalent ways in order to facilitate the comparison with the literature.\footnote{The form (\ref{spectrum}) is simpler and more natural but (\ref{spectrum0}) has the same form as the scalar QNM solution \cite{Cardoso:2001hn} that we want to benchmark against.}

Another way to derive the QNM spectrum is by definition, as the eigenfrequencies of the equations of motion with infalling boundary conditions at the horizon and Dirichlet boundary conditions at the boundary. The latter require the solution at AdS boundary to vanish.\footnote{In this context we again ignore the asymptotically flat region of the D1-D5-p geometry and only consider its interior and AdS throat, just as we did when solving the string equations of motion. In an asymptotically Minkowski spacetime we would require outgoing boundary condition at infinity, very different from the situation in AdS. This will be important in what follows.} The equivalence of the two approaches should be obvious, since the same set of requirements that force the solution (\ref{2F1exp}) to vanish at infinity also describes the poles of the retarded Green's function (\ref{2ptcorr}). This gives us a more intuitive picture of QNM: they tell us how a local near-horizon instability decays. Therefore, we can think of the inverse of the Lyapunov exponent $\lambda_L^{-1}$ as some characteristic timescale for the decay of perturbations along the open string, that has nothing to do with chaos. 

The result summarized in Eqs.~(\ref{spectrum0}-\ref{spectrum}) is qualitatively the same as the one obtained in \cite{Cardoso:2001hn} for scalar perturbations in a nonrotating BTZ black hole background, except that the spectrum (\ref{spectrum}) also includes half-integer values of $\mathfrak n$ (which is simply due to different objects being considered: strings vs. scalar field). Similar scaling of the QNM spectrum with the Lyapunov exponent of some special orbit is known from two classic papers concerning \emph{asymptotically planar} black holes:
\begin{enumerate}
\item In \cite{Cardoso:2008bp}, the imaginary part of the QNM frequencies is determined by the instability of geodesic motion, i.e. the Lyapunov exponent of a massless particle \emph{on the photon sphere}, which acts as an unstable fixed point. This is obtained in the eikonal approximation when $\Re \omega \gg \Im \omega$.\footnote{Here $\Re \omega$ denotes the orbital frequency (oscillations) and $\Im \omega$ is the Lyapunov exponent for the unstable geodesic orbit (damping).}
\item In the opposite, high-overtone or overdamped regime, when $\mathfrak n \gg 1$ and thus $\Re \omega \ll \Im \omega$, \cite{Motl:2003cd} find that $\Im\omega$ is proportional to the surface gravity at the horizon, which equals precisely $2\pi T$.
\end{enumerate}
Our results are clearly obtained in the high-overtone regime since our Green's functions satisfy $\Im \mathcal G_R^{(T)} \gg \Re \mathcal G_R^{(T)}$. Therefore, our result essentially generalizes \cite{Motl:2003cd} for an open string in AdS. Yet, the coefficient itself is obtained as the Lyapunov exponent of an unstable fixed point, and in that sense generalizes also \cite{Cardoso:2008bp} from the eikonal (photon-sphere dominated) regime to the high-overtone (horizon-dominated) regime.

We have thus shown that both "easy" regimes ($l\gg 1$ and $\frak{n}\gg 1$) can be understood from classical unstable saddle points, but of course the eikonal regime sees the scattering near the photon sphere while the overdamped regime (our case) sees the horizon. In this sense, earlier results on the universal $2\pi T$ exponent for geodesics and fields near-horizon \cite{Giataganas:2021ghs}, bringing the conclusion about the horizon as the "nest of chaos", are not in collision with the studies of the instability on the photon sphere \cite{Cardoso:2008bp,Bianchi:2020yzr,Bianchi:2022mhs} -- only that they correspond to different regimes. It is somewhat surprising that for a special string configuration these results are obtained in asymptotically AdS backgrounds, as it is well known and discussed already in \cite{Cardoso:2008bp,Motl:2003cd} that for a geodesic the argument does not easily generalize to AdS asymptotics.

Finally, we should also comment on the field theory interpretation of the quasinormal modes spectrum that we have just found in the bulk. We already mentioned that an open string in the bulk stretched from the boundary to the thermal AdSS horizon corresponds to a heavy quark in thermal plasma of super-Yang-Mills quarks and gluons. Perturbations along the string describe thermal perturbations in the plasma. Similar holds in the D1-D5 CFT except that the elementary excitations now cannot be called quarks. We can summarize the findings above by noting that a Lyapunov exponent is really related to the QNM frequencies, which describe how local near-horizon instabilities on the string decay. Decay rates of those instabilities are given by the spectrum of quasinormal modes, so on the field theory side they describe how the thermal fluctuations in plasma die off. Thus, they predict the thermalization timescale in the dual CFT.

\section{Discussion and conclusions}

The initial motivation for this work was a rather technical question: what is the meaning of bulk chaos in particle and string motion in AdS spaces, and why it typically saturates the same universal chaos bound as OTOC in field theory. We were led to the study of open strings (rather than ring strings or particle geodesics) largely by reasons of calculational simplicity and direct CFT interpretation: a string with one end on the boundary and the other in the interior describes the motion of a heavy quark in quark-gluon plasma. The holographic interpretation is less obvious for other string configurations, and for geodesics it corresponds to a rather special, high-conformal-dimension limit.

As usual, the chase is almost better than the catch. We have found a number of surprising properties of bulk dynamics, first and foremost the horizon as an unstable saddle point and the "fake nest of chaos" with local instability rate exactly equal to $2\pi T$ in the static case but different from it in the rotating black string geometry. But the holographic interpretation is equally interesting: the universal MSS-like exponent is a red herring, the artifact of the large-$N$ limit in field theory, i.e. classical bulk dynamics, when temperature is the only scale, unless some additional symmetry is explicitly broken in IR. This happens in the D1-D5-p black string, where the rotation breaks the symmetry between left- and right-moving modes. Just like in \cite{Craps:2021bmz}, the rotating system deforms away from the universal $2\pi T$ exponent, and this shows directly in the correlation functions. We also note that away from the dilute gas approximation there are additional higher-order temperature corrections to the Lyapunov exponent  -- this is the effect of the UV deformation.

We have found the connection between the near-horizon Lyapunov exponent and the spectrum of quasinormal modes in the high-overtone regime. This gives us an important hint about the meaning of the bulk Lyapunov exponent: it is an instability scale associated to the decay of fluctuations along the string due to thermal dissipation, and has nothing to do with bulk chaos. This resembles two classic results on quasinormal modes of asymptotically flat black holes: in the eikonal regime the Lyapunov exponent on the photon sphere determines the quasinormal mode \cite{Cardoso:2008bp}, whereas in the overdamped regime (our case) the imaginary part of the quasinormal mode equals the horizon surface gravity \cite{Motl:2003cd}. We have essentially shown that for near-horizon string orbits the bulk Lyapunov exponent equals the surface gravity (which generalizes even for a rotating horizon).

Motivated by this connection, one could try to extend some other known results on geodesic instabilities to the orbits of extended objects like strings. Of primary interest is the instability on the photon sphere which, apart from the well-known connection with quasinormal modes, holds the key to several other properties both in asymptotically flat and in global AdS spaces \cite{Hashimoto:2023buz,Riojas:2023pew}. However, our setup needs to be substantially modified to study the photon sphere, which arises as the locus of unstable saddle points for null geodesics \emph{arriving from infinity}. An open string in our configuration does not even have a saddle point on the photon sphere, i.e. one cannot even define the Lyapunov exponent on the photon sphere unambiguously. Therefore, instead of having a static string, we would need to scatter an open string along a null geodesic. In that case one would expect the minimal allowed value for the impact parameter to be of order of the photon sphere size (analogous study for massless particles is performed in \cite{Bianchi:2020des}). Such a setup is particularly suitable for the study of fuzzballs and microstate geometries, since they have no sharp length/energy scale analogous to a horizon \cite{Bianchi:2020yzr,Bianchi:2021yqs}. We would then be probing another branch of the QNMs spectrum, associated to stringy instabilities around the photon sphere.

One might find it surprising that our open string lives in an integrable sector. This is likely a consequence of the highly symmetric and simple boundary conditions for which we prove integrability: a static string at the horizon. It is known that a ring string is nonintegrable in thermal backgrounds, and also in generic Dp-brane backgrounds (although some very special cases can be integrable, even at finite temperature, see e.g. \cite{Roychowdhury:2020zer,Pal:2023bjz}). A more generic open string configuration is likely also nonintegrable. Essentially, since \emph{different boundary conditions for string motion result in a different effective Lagrangian, the Liouville integrability has to be checked separately for every configuration}. While \emph{physically} we like to think of "string motion in a given background", mathematically the system is integrable if the Euler-Lagrange equations satisfy certain conditions -- and for a string the form of these equations depends on the string ansatz. Therefore, while a single example is enough to prove general nonintegrability, proving integrability in the most general case requires more powerful tools than the Kovacic algorithm. We do not do that: we merely focus on a single case which turns out to be integrable.

From our results it is clear that the study chaos and scrambling in gauge/string duality (i.e., beyond classical gravity) is a separate topic, not much touched upon in this work. Fluctuations of a static straight string stretched from boundary to boundary of a maximally extended (static and neutral) BTZ black hole are known to lead to the exact MSS value for the worldsheet OTOC \cite{DeBoerVeghStringBnd}; for a dragging string, butterfly velocity also enters the picture, as the leading correction to the drag force \cite{Ageev:2021poy}. Given the high symmetry of these systems, this is not surprising; when the symmetry is decreased or corrections added to the classical string solution the value will be modified but there is no reason to believe that the solutions and the OTOC exponents will be modified in the same way as the bulk exponents, i.e. quasinormal modes in our setup.  Worldsheet scrambling for a rotating BTZ black hole was studied in \cite{Banerjee:2019vff} and indeed, while the OTOC exponent is modified from the MSS value, it is not the same as the bulk exponent that we find here. Of course, strings can also model spatiotemporal chaos if we allows both worldsheet coordinates to fluctuate as in \cite{Ishii:2023qqd}.

Finally, the issue of gauge choice might be worth commenting. Our choice to work in the conformal gauge instead of static gauge most of the time is somewhat unusual. The static gauge equates the time and radial coordinate with the worldsheet coordinates $\tau$ and $\sigma$ and thus immediately kills the unphysical (gauge-dependent) degrees of freedom. But the conformal gauge has several advantages for us: (\textit{i}) it simplifies many calculations (\textit{ii}) it allows us to look at the fluctuations along the holographic RG flow (the radial direction) (\textit{iii}) it does not fully fix the reparametrization invariance on the worldsheet, leaving the $\mathrm{SL}(2,\mathbb{R})$ group of global coordinate transformations, but as argued in \cite{Lin:2019qwu,fake:geometry,Giombi:2022pas} this group provides a nice way to understand the appearance of a universal scale and its disappearance when we determine the boundary conditions for the transverse fluctuations that fully fix the gauge on the worldsheet. This approach was exploited in full depth in \cite{Giombi:2022pas} to study quantum chaos, i.e. OTOC on the worldsheet of the open string.

\subsection{Note added: quasi-normal modes, variational equations and the spectral form factor}

At the end we want to comment on another connection between our calculation and the quasi-normal modes of the black hole (or black string) background. After finishing the first version of the paper we became aware of the work \cite{Chen:2023hra} where it is shown that the partition function and the spectral form factor of a holographic theory at finite temperature can be understood as a product over the QNM frequencies $\omega_{\mathrm{QNM}}$ of the bulk black hole. Specifically, for a bosonic system at temperature $T$, one-loop partition function is found in \cite{Chen:2023hra} to be
\begin{equation}
Z=\mathrm{Tr}e^{-i\hat{\tilde{K}}T}=\prod_{\omega_{\mathrm{QNM}}}\left(1-e^{-i\omega_{\mathrm{QNM}}T}\right)^{-1}.\label{qnmeq}
\end{equation} 
Here, $\hat{\tilde{K}}$ is the time-shift operator which acts as a boost near the horizon (where $g_{tt}\to 0$), in a complexified metric where the radial coordinate is shifted as $r\mapsto r-i\epsilon$. From (\ref{qnmeq}), it is obvious that $\omega_{\mathrm{QNM}}$ are just the eigenvalues of $\hat{\tilde{K}}$. But Lyapunov exponents are nothing but the eigenvalues of the Jacobian matrix of the equations of motion -- in other words, they are the eigenvalues of the shift operator but now the shift is along the tangential directions in phase space. In a given background however, e.g. in a near-horizon region like the BTZ region of a near-extremal black string, one can choose the gauge so that the Jacobian \emph{locally} (but not everywhere) coincides with the time-shift operator $\hat{K}$; the complexification to $\hat{\tilde{K}}$ then just imposes the analytic behavior at the horizon, as one normally does when computing correlation functions such as $G_R$ from subsection \ref{subsecd1d5trans}.

The above discussion is obviously not rigorous. It would be interesting to formulate it in strict terms and see how much one can learn from such a viewpoint.

\acknowledgments

We are grateful to Andrei Parnachev, Jorge Russo, David Berenstein, Filip Her\v{c}ek and Juan Pedraza for stimulating discussions. Work at the Institute of Physics is funded by the Ministry of Education, Science and Technological Development and by the Science Fund of the Republic of Serbia. M.~\v{C}. would like to acknowledge the Mainz Institute for Theoretical Physics (MITP) of the Cluster of Excellence PRISMA+ (Project ID 39083149) for hospitality and partial support during the completion of this work.

\appendix

\section{Slightly generalized ansatz for the Dp-brane background}\label{secappa}

Here we comment on the ansatz for the string configuration in Dp-brane backgrounds in subsection \ref{subsecdbrane}. We can have nontrivial dynamics of the string also on the $k$-sphere provided we impose some additional constraints which are necessary to preserve the separation of variables. For example, we can replace the ansatz from Eq.~(\ref{dbranestringconfig}) by
\begin{eqnarray} \nonumber
    t&=&t(\tau), \quad X_1 = x_1, \quad \ldots \quad X_{10-k}=x_{10-k}, \\
    R&=&R(\sigma), \quad \Phi_1 = \Phi_1(\tau),\quad \Phi_2 = \Phi_2(\tau),\quad \Phi_3 = \phi_3,\quad \ldots\quad \Phi_k = \phi_k\label{dbranestringconfig2}
\end{eqnarray}
Choosing $t(\tau) = \tau$ as usual, we are left with a constraint (in addition to the Virasoro constraint) coming from the above ansatz, i.e. the assumption that $R$ only depends on $\sigma$:
\be
\dot{\Phi}_1^2+\sin\Phi_1^2\dot{\Phi}_2^2\equiv \ell^2, \label{ell}
\ee
where $\ell^2$ is the conserved squared angular momentum on the $k$-sphere. The constraints decouple the dynamics of $R$ from $\Phi^1$ and $\Phi^2$, so the equation of motion for $R$ remains the same as Eq.~(\ref{dbraneeomr}) and for $\Phi_1$ we obtain:
\begin{equation}
\ddot{\Phi}_1+\left(\dot{\Phi}_1^2-w^2\right)\cot\Phi_1=0.\label{dbraneeomphi}
\end{equation}
Therefore, it is possible to go for more general dynamics than in the main text, which might be of interest for some applications but is completely peripheral for our main interest in this paper.

\bibliography{stringsbibopen.bib}

\end{document}